# Thermonuclear Processes for Three Body System in the Potential Cluster Model

*S.B. Dubovichenko, A.V. Dzhazairov-Kakhramanov*

V. G. Fessenkov Astrophysical Institute "NCSRT" RK

050020, Kamenskoe plato 23, Observatory, Almaty, Kazakhstan

dubovichenko@mail.ru, albert-j@yandex.ru

**Abstract** The manuscript is devoted to the description of the results obtained in the frame of the modified potential cluster model with the classification of states according to Young tableaux for neutron and proton radiative capture processes on $^2$H at thermal and astrophysical energies. It demonstrates methods of application that were obtained on the basis of phase shift analysis and characteristics of the bound states of $^2$H potentials for consideration of the radiative capture processes. First reaction of the proton capture directly takes part in the pp solar cycle, where it is the second reaction. The neutron capture is not a part of usual thermonuclear cycles in the Sun and stars, but can take part in the processes of primordial nucleosynthesis, following at formation and evolution of our entire Universe.

## 1 Introduction

One extremely successful line of development of nuclear physics in the last 50-60 years has been the microscopic model known as the Resonating Group Method (RGM, see, for example [1-4]). And the associated with it models, for example, Generator Coordinate Method (see, particularly, [4]) or algebraic version of RGM [5]. However, the rather difficult RGM calculations are not the only way in which to explain the available experimental facts.

But, the possibilities offered by a simple two-body potential cluster model (PCM) have not been studied fully up to now, particularly if it uses the concept of forbidden states (FSs) [6]. The potentials of this model for discrete spectrum are constructed in order to correctly reproduce the main characteristics of the bound states (BSs) of light nuclei in cluster channels, and in the continuous spectrum they directly take into account the resonance behavior of the elastic scattering phase shifts of the interactive particles at low energies [7].

As it was shown in work [7], it is enough to use the simple PCM with FSs taking into account the classification of the orbital states according to Young tableaux and resonance behavior of the elastic scattering phase shifts – such a model can be called as a modified PCM (MPCM). In many cases, such an approach, as has been shown previously, allows one to obtain adequate results in the description of many experimental studies for the total cross sections of the thermonuclear reactions at low and astrophysical energies [7].

Particularly, in works [8], we have sown the possibility of description the Coulomb form-factors of lithium nuclei on the basis of potential cluster model [7,9]. As we have just said, this model takes into account forbidden states [9-12] in the intercluster potentials, which are determined on the basis of the classification according Young tableaux and were used by us in works [13]. Furthermore in works [14] we show the possibility of the correct reproduction practically all characteristics of $^6$Li, including is quadrupole moment in the potential cluster model with tensor forces [10]. And finally, in works [7,9,15-20] the possibility of description of the astrophysical *S*-factors or the total cross sections of the radiative capture for $n^2$H, $p^2$H, $p^3$H, $n^6$Li, $p^6$Li, $n^7$Li, $p^7$Li, $p^9$Be, $n^9$Be, $p^{10}$B, $n^{10}$B, $p^{11}$B, $n^{11}$B, $n^{12}$C, $p^{12}$C, $n^{13}$C, $p^{13}$C, $n^{14}$C, $p^{14}$C, $n^{14}$N, $n^{15}$N, $p^{15}$N, $n^{16}$O and $^2$H$^4$He, $^3$He$^4$He, $^3$H$^4$He, $^4$He$^{12}$C systems at thermal and astrophysical energies. These calculations of the listed above capture processes are carried out on the basis of the modified variant of PCM, described in works [7,19,21].

Therefore, continuing consideration of thermonuclear processes flowing in the different objects of the Universe at the different stage of its formation and development [7], more particularly present some results in the research field of these reaction at thermal and astrophysical energies. New results in the research area of thermonuclear reactions at lowest, thermal, and astrophysical energies are presented here. The two-body MPCM model is used as the nuclear model, which allows us to consider some thermonuclear processes, notably, reactions of the proton and neutron radiative capture on the basis of unified conceptions, criteria, and methods. Furthermore we have considered the total cross sections of two proton and neutron radiative capture processes on $^2$H. It is shown that the classification of the cluster state according to Young tableaux gives the possibility to construct potentials of the continuous and the discrete spectra, which allow us to describe these processes and some basic static characteristics of nuclei $^3$H and $^3$He. In the first section – the brief description of the model and its methods is given, in the second – the results for the proton radiative capture on $^2$H, and in the third – the neutron capture is considered on the same nucleus.

## 2 Model and methods

Many characteristics of light atomic nuclei, which take part in thermonuclear processes, can be well described by different variants of nuclear models and one of them is the potential cluster model [6,7,22]. We are using the modified variant of such cluster model of light atomic nuclei with the classification of the orbital states according to Young tableaux [6,8,22]. This model gives comparatively much easy possibilities for carrying out different calculations of astrophysical characteristics, for example, the astrophysical *S*-factor or total cross sections of radiative capture for electromagnetic transitions from the scattering cluster states to the bound states of light atomic nuclei in these cluster channels [7].

### 2.1 Basic principles of the model

The efficiency of the MPCM is determined by the fact that, in many light atomic nuclei, the probability of formation of nucleon associations (clusters) and the degree of separation from each other are rather high. This is proved by multiple experimental data and theoretical calculations obtained during the last 50-60 years [6]. For construction of phenomenological potentials of intercluster interaction, the results of phase shift analysis of experimental data on differential cross sections of elastic scattering of corresponding free nuclei are used [8]. Potentials of the scattering processes are constructed from the condition of the best description of the obtained elastic scattering phase shifts, which were obtained on the basis of these data. The potentials of the bound state (BS) of clusters are constructed, as a rule, on the basis of description of certain characteristics of the ground state (GS) of nuclei, which is considered in this cluster channel.

However, the results of phase shift analysis in the limited energy range, as a rule, prevent unambiguous reconstruction of the interaction potential. Therefore, the additional restriction for the intercluster potential is the requirement of its agreement with the results of classification according to Young tableaux, i.e., instead of allowed states (ASs), if they exist, it has to include, as a rule, certain number of FSs. At the construction of the potential of the ground or excited, but bound in the considered channel state, is the additional condition of reproduction of the binding energy of the nucleus in the corresponding cluster channel and some other static nuclear properties. It may be, for example, charge radius and asymptotic constant (AC), meanwhile the characteristics of the binding clusters in nuclei are identified with characteristics of corresponding free lightest nuclei [18]. This additional requirement, obviously, is an idealization, since it assumes that the BS of the nucleus is 100% clusterized. Actually, the success of this potential model, in description of a system of $A$ nucleons in the bound state, is determined by the actual degree of clusterization of the ground state of such nucleus in two- and three-body channels [6,18,22]. However, some nuclear characteristics of particular, even non-cluster, nuclei can be mainly determined by one specific cluster channel and the small contribution of other possible cluster configurations. In this case, the applied single-channel cluster model makes it possible to identify the dominating cluster channel and separate those properties of the cluster system that are determined by this channel [18].

The considered potential cluster model is rather simple in application, since technically it comes to solution of the two-body problem, or, which is equivalent, to the problem of one body in the field of a force center. Therefore, an objection can be put forward that this model is absolutely inadequate to the many-body problem to which the problem of description of properties of the system consisting of $A$ nucleons is related. In this regard, it should be noted that one of the successful models in the theory of atomic nucleus is the model of nuclear shells (SM) that mathematically represents the problem of one body in the field of a force center. The physical grounds of the potential cluster model considered here trace to the shell model or, more precisely, in a surprising connection between the shell model and the cluster model, which is mentioned in the literature as the nucleon association model (NAM) [6,18].

In the NAM and PCM, the wave function (WF) of the nucleus consisting of two clusters with the numbers of nucleons $A_1$ and $A_2$ ($A = A_1 + A_2$) has the form of antisymmetrized product of totally antisymmetric internal wave functions of clusters $\Psi(1, …, A_1) = \Psi(R_1)$ and $\Psi(A_1+1,…,A) = \Psi(R_2)$ multiplied by the wave function of their relative motion $\Phi(R = R_1 - R_2)$,

$$\Psi = \hat{A} \{\Psi(R_1)\Psi(R_2)\Phi(R)\}, \qquad (1)$$

where $\hat{A}$ is the operator of antisymmetrization under permutations of nucleons belonging to different clusters, $R$ is the intercluster distance, $R_1$ and $R_2$ are the radius vectors of the position center mass of clusters.

Usually cluster wave functions are chosen in such a way that they correspond to ground states of nuclei consisting of $A_1$ and $A_2$ nucleons. These shell wave functions are characterized by specific quantum numbers, including Young tableaux $\{f\}$, which determine the permutation symmetry of the orbital part of cluster relative motion WF. In addition, certain conclusions of the cluster model [6,18] lead to the concept of Pauli-forbidden states. Therefore, some total WFs of nucleus $\Psi(R)$ with the certain type of relative motion functions $\Phi(R)$ go to zero at the antisymmetrization by whole $A$ nucleons (1).

Ground, i.e., really existed bound state of the cluster system, in this potential, is described by the wave function with nonzero, in general, number of nodes. Thereby, the conception about Pauli-forbidden states allows one to take into account the multi-body character of the problem in terms of two-body interaction potential between clusters [6,18]. Meanwhile, in practice, the potential of the intercluster interaction is chosen so that to correctly describe the cluster scattering phase shifts extracted from the experimental data of corresponding partial wave and, preferentially, in the state with one certain Young tableau $\{f\}$ for spatial part of the wave function of $A$ nucleons of the nucleus [6,22].

## 2.2 Potentials and wave functions

Intercluster interaction potentials for each partial wave, i.e., for the given orbital angular moment $L$, and point-like Coulomb term, were represented as

$$V(r,L) = V_0(L)\exp(-\alpha_L r^2) + V_1(L)\exp(-\delta_L r) \qquad (2)$$

or

$$V(r,L) = V_0(L)\exp(-\alpha_L r^2). \qquad (3)$$

Here, parameters $V_0$ and $V_1$, $\alpha$ and $\gamma$ are the potential parameters, which, for example, are found from experimental data under the constraint of the best description of elastic scattering phase shifts extracted in the course of phase shift analysis from the experimental data on the differential cross sections, i.e., angular distributions or excitation functions and can contain FSs.

The Coulomb potential includes the Coulomb radius $R_{\text{Coul}} = 0$ and then the Coulomb potential takes the form

$$V_{\text{Coul}}(\text{MeV}) = 1.439975\, Z_1 Z_2 / R, \qquad (4)$$

where $r$ is the relative distance between particles of the initial channel in fm and $Z$ are charges of particles in the elementary unit charge "$e$".

The behavior of the wave function of bound states, including ground states of nuclei in cluster channels at large distances, is characterized by the asymptotic constant $C_W$ determining by the Whittaker constant of the form [23]

$$\chi_L(r) = \sqrt{2k_0}\, C_W W_{-\eta L+1/2}(2k_0 r), \qquad (5)$$

where $\chi_L(R)$ is the numerical wave function of the bound state obtained from the solution of the radial Schrödinger equation and normalized to unity; $W_{-\eta L+1/2}$ is the Whittaker function of the bound state determining the asymptotic behavior of the wave function which is the solution to the same equation without the nuclear potential, i.e., at large distances $R$; $k_0$ is the wave number determined by the channel binding energy; $\eta$ is the Coulomb parameter determined further; and $L$ is the orbital angular moment of the bound state.

Asymptotic constant (AC or as it is often called asymptotic normalization coefficient – ANC) is an important nuclear characteristic determining behavior of the "tail", i.e., asymptotic of the wave function at the large distances. In many cases the knowledge of its value for the $A$ nucleus in the $b + c$ channel determines the value of the astrophysical $S$-factor of the radiative $b(c, \gamma)A$ capture process [24]. The asymptotic constant is proportional to the nuclear vertex constant for the virtual $A \rightarrow b + c$ process, which is the matrix element of this process at the mass surface [25].

The numerical wave function $\chi_L(R)$ of the relative motion of two clusters is the solution of the radial Schrödinger equation of the form

$$\chi''_L(r) + [k^2 - V_n(r) - V_{Coul}(r) - L(L+1)/r^2]\chi_L(r) = 0, \tag{6}$$

where $V_{Coul}(r) = 2\mu/\hbar^2 Z_1 Z_2/r$ is the Coulomb potential leading to the dimension fm$^{-2}$, $Z_1$ and $Z_2$ are the particle charges in units of elementary charge, $k^2 = 2\mu \dfrac{m_0}{\hbar^2} E$ is the wave number of particle relative motion in fm$^{-2}$, $E$ is the energy of particles, $\mu = m_1 m_2/(m_1+m_2)$ is the reduced mass of two particles, $V_n(r)$ is the nuclear potential equals $2\mu/\hbar^2 V(r)$, $V(r)$ is the radial dependence of the potential frequently takes the forms (2) or (3) and leading to the dimension fm$^{-2}$, the constant $\hbar^2/m_0$ was equal to 41.4686 MeV fm$^2$, $m_0$ is the atomic mass unit (amu). Although this value, as of today, is considered slightly out-of-date, but we continue to use it for relieving of comparison the last and the earlier obtained results (see, for example, [7,8,18]). The Coulomb parameter $\eta = \dfrac{\mu Z_1 Z_2 e^2}{\hbar^2 k}$ was represented as [26]

$$\eta = 3.44476 \cdot 10^{-2} \dfrac{\mu Z_1 Z_2}{k}, \tag{7}$$

where $k$ is the wave number $k = \sqrt{2\mu \dfrac{m_0}{\hbar^2} E}$, leading to the dimension fm$^{-1}$, $\mu$ is the reduced mass, $Z_{1,2}$ are the particle charges in units of elementary charge.

Asymptotics of the scattering wave function $\chi_L(r)$ at large distances $R \rightarrow \infty$, etc. at $V_n(r \rightarrow R) = 0$ is the solution of the equation (6) and can be presented in the next form

$$\chi_L(r \rightarrow R) \rightarrow F_L(kr) + \text{tg}(\delta_L)G_L(kr) \tag{8}$$

or

$$\chi_L(r \to R) \to \cos(\delta_L)F_L(kr) + \sin(\delta_L)G_L(kr), \quad (9)$$

where $F_L$ and $G_L$ are the wave Coulomb scattering functions [27], which are the particular solutions of the equation (6) without nuclear part of the potential, i.e., when $V_n(r) = 0$.

The numbering solution $\chi_L$ of the equation (6) matches the asymptotics scattering processes at distances about 10–20 fm, and it allows one to find the scattering phases $\delta_{LJ}$ for each value of orbital moments $L$ at the given energy of interacting particles. The scattering phase shifts in the specific system of nuclear particles can be obtained from the phase shift analysis of the elastic scattering experimental data. Furthermore, the variation of parameters of the nuclear potential of the previously defined forms of equation (6) will be done, and the parameters that allow us to describe the results of phase shift analysis are determined. Thereby, the problem of description of the scattering processes of nuclear particles is in the search of parameters of nuclear potential, which describe the results of phase shift analysis, and so as the experimental data on the scattering cross section.

## 2.3 Generalized matrix eigenvalues problem

Considering the generalized matrix eigenvalues and eigenfunctions problem that is the result of WF expansion on the nonorthogonal Gaussian basis

$$\Phi_L(r) = \frac{\chi_L(r)}{r} = N_0 r^L \sum_i C_i \exp(-\beta_i r^2), \quad (10)$$

we come from the standard Schrödinger equation in the general form [28]

$$H\Phi = E\Phi, \quad (11)$$

where $H$ is the Hamiltonian of a system, $E$ is the energy of system and $\Phi$ wave functions of the relative motion of two particles, $N_0$ is the normalization coefficient.

Expand the WF into certain, nonorthogonal in the general case, variational basis [29]

$$\Phi = \sum_i C_i \varphi_i, \quad (12)$$

and substituting them into the initial system, product it from the left on the complex conjugate basis function $\varphi_i^*$ and integrate it by all variables, we will obtain known matrix system of the form [7]

$$(\mathbf{H} - E\mathbf{L})C = 0, \quad (13)$$

which, in the general case, is the generalized matrix problem for finding eigenvalues and eigenfunctions [7]. If the expansion of the WF is done according to orthogonal basis, the matrix of overlap integrals $\mathbf{L}$ turn into the unity matrix $\mathbf{I}$, and we have standard problem on eigenvalues and for solving it there are a lot of methods [7].

There are known methods for solution of the generalized matrix problem, for example, given in book [7]. Let us stop, firstly, on the standard method of solution of the Schrödinger equation, which appears at using the nonorthogonal variational basis in nuclear physics and nuclear astrophysics. Then, consider its modification or alternative method, which will be convenient for solving this problem by numerical calculations using up-to-date computers [7].

Thus, the generalized matrix problem on eigenvalues is solved at the determination of the spectrum of energy eigenvalues and eigen wave functions in the variational method, at the expansion of the WF according to nonorthogonal Gaussian basis [29-31].

$$\sum_i (\mathbf{H}_{ij} - E\mathbf{L}_{ij})C_i, \qquad (14)$$

where **H** is the symmetric Hamiltonian matrix; **L** is the matrix of overlapping integrals; $E$ are the energy eigenvalues; and $C$ are the eigenvectors of the problem.

Representing the matrix **L** in the form of the product of the lower **N** and upper **V** triangular matrices [7], after simple transformations, we obtain the common eigenvalues problem

$$\mathbf{H'}C' = E\mathbf{I}C', \qquad (15)$$

or

$$(\mathbf{H'} - E\mathbf{I})C' = 0, \qquad (16)$$

where

$$\mathbf{H'} = \mathbf{N}^{-1}\mathbf{H}\mathbf{V}^{-1}, \ C' = \mathbf{V}C, \qquad (17)$$

where $\mathbf{V}^{-1}$ and $\mathbf{N}^{-1}$ are inverse to the **V** and **N** matrices, respectively.

Furthermore, we find the matrices **N** and **V**, performing triangularization of the symmetric matrix **L** [7], for example, using the Khaletskii method [7]. Then we determine the inverse matrices $\mathbf{N}^{-1}$ and $\mathbf{V}^{-1}$, for example, using the Gauss method [7], and calculate the elements of the matrix $\mathbf{H'} = \mathbf{N}^{-1}\mathbf{H}\mathbf{V}^{-1}$. We find the complete diagonal with respect to the $E$ matrix $(\mathbf{H'} - E\mathbf{I})$ and calculate its determinant $\det(\mathbf{H'} - E\mathbf{I})$ for some energy $E$. The energy resulting in the zero determinant is the eigenenergy of the problem, and the corresponding vectors $C'$ are the eigenvectors (15). If $C'$ are known, it is easy to find the eigenvectors of the initial problem $C$ (13), since the matrix $\mathbf{V}^{-1}$ is already known. The described method of reduction of the generalized matrix problem to the common matrix problem is called the Schmidt orthogonalization method [32].

However in some numerical problems at certain values of variational parameters $\beta_i$ the procedure of finding reverse matrices appeared to be unstable and during the work of computer program the overflow is done [7]. Therefore, an alternative method for numerical solution of the generalized matrix eigenvalues problem free from the difficulties indicated above with enhanced computer performance can be proposed. That is to say that initial matrix Eqs. (13) or (14) are the homogeneous system of linear equations and has nontrivial solutions only if its determinant $\det(\mathbf{H} - E\mathbf{L})$ is equal to zero. For computer numerical methods, it is not necessary to expand the matrix **L** into triangular matrices and find the new matrix **H'** and new vectors **C'** by determining inverse matrices, as was described above using the standard method. It is possible to expand the nondiagonal symmetric matrix (**H** - $E$**L**) into triangular matrices and seek energies resulting in zero determinant, i.e., eigen energies, using numerical methods in the given domain. In the real physical problem, usually, it is not necessary to search all eigenvalues and eigenfunctions. It is necessary to find only one or two eigenvalues for certain energy of the system and, as a rule they are the lowest values and corresponding to them eigen wave functions.

Therefore, the initial matrix (**H** - $E$**L**) can be expanded into two triangular matrices using, for example, the Khaletskii method, in such a way that the main diagonal of the upper triangular matrix **V** contains units,

$$\mathbf{A} = \mathbf{H} - E\mathbf{L} = \mathbf{NV} \tag{18}$$

the determinant of this matrix for det(**V**) = 1 is calculated [7],

$$D(E) = \det(\mathbf{A}) = \det(\mathbf{N}) \det(\mathbf{V}) = \det(\mathbf{N}) = \prod_{i=1}^{m} n_{ii} \tag{19}$$

and the zero of this determinant is used to find the required energy eigenvalue, i.e., the value $E$. Here, $m$ is the dimensionality of the matrices and the determinant of the triangular matrix **N** is equal to the product of its diagonal elements [7].

Thus, we obtain a rather simple problem of finding the zero of a functional of one variable,

$$D(E) = 0, \tag{20}$$

numerical solution of this problem does not present great difficulty and can be found with any accuracy, for example, using division into halves.

As a result we eliminate the necessity of finding both inverse to **V** and **N** matrices and carry out several matrix multiplications in order to first obtain the new matrix **H'** and then the final matrix of eigenvectors **C**. The absence of such operations, especially finding of inverse matrices, leads to computer counting rate increasing independently of code languages that we use for solving the this problem [7].

For estimation of the solution accuracy, i.e., the accuracy of expansion of the initial matrix into two triangular matrices, the notation of residuals for matrix elements was used. After expansion the matrix **A** into two triangular matrices, the residue matrix [7] is calculated as the difference between initial matrix **A** and matrix

$$\mathbf{S} = \mathbf{NV}, \tag{21}$$

where **V** and **N** are the found numerical triangular matrices. Now the difference up to all components with the initial matrix **A** is taken

$$\mathbf{A_N} = \mathbf{A} - \mathbf{S}. \tag{22}$$

The residue matrix $\mathbf{A_N}$ gives the deviation of the approximate value **NV**, found by the numerical methods, from the true value of each element of the initial matrix **A**. One can carry out summation of all matrix elements **A**$_N$ and obtain the numerical value of the residual δ.

This method, which seems quite obvious in numerical implementation, made it possible to obtain good stability of the algorithm for solution of the considered problem; it does not result in overflow in the course of running the computer program [7]. The described here method was used in all carried out variational calculations [7] and the maximal value of any matrix element **A**$_N$ usually is not more than $10^{-10}$. Hereby, the introduced alternative method of finding eigenvalues of the generalized matrix problem, considered on the basis of the Schrödinger equation solution using the non-orthogonal variational basis, delivers us from the instabilities appearing during the use of normal solution methods of such mathematical problem, i.e., usual Schmidt orthogonalization method.

## 2.4 Total radiative capture cross sections

The total radiative capture cross sections σ(*NJ*,*J*$_f$) for the *EJ* and *MJ* transitions in the potential cluster model are given, for example, in [7,8] or [33] and are written as

$$\sigma_c(NJ,J_f) = \frac{8\pi K e^2}{\hbar^2 q^3} \frac{\mu}{(2S_1+1)(2S_2+1)} \frac{J+1}{J[(2J+1)!!]^2} A_J^2(NJ,K) \cdot$$

$$\cdot \sum_{L_i,J_i} P_J^2(NJ,J_f,J_i) I_J^2(J_f,J_i), \qquad (23)$$

where σ is the total cross section of the radiative capture process, μ is the reduced mass of particles in the initial channel, *q* is the wave number of particles in the initial channel, $S_1$ and $S_2$ are the spins in the initial channel, *K* and *J* are the wave number and angular moment of γ-quantum in the final channel, *N* these are *E* or *M* transitions with multipole order *J* from the initial $J_i$ to the final $J_f$ state of the nucleus.

The $P_J$ value for electric orbital *EJ*(*L*) transitions ($S_i = S_f = S$) has the form [7,8]:

$$P_J^2(EJ,J_f,J_i) = \delta_{S_iS_f}[(2J+1)(2L_i+1)(2J_i+1)(2J_f+1)](L_i 0 J 0 | L_f 0)^2 \begin{Bmatrix} L_i & S & J_i \\ J_f & J & L_f \end{Bmatrix}^2$$

$$A_J(EJ,K) = K^J \mu^J \left(\frac{Z_1}{m_1^J} + (-1)^J \frac{Z_2}{m_2^J}\right), \qquad I_J(J_f,J_i) = \langle \chi_f | R^J | \chi_i \rangle \qquad (24)$$

Here $S_i$, $S_f$, $L_f$, $L_i$, $J_f$, $J_i$ are the total spins and angular moments of particles in the initial (*i*) and the final (*f*) channels, $m_1$, $m_2$, $Z_1$, $Z_2$ are masses and charges of the particles in the initial channel, respectively, $I_J$ is the integral over wave functions of the initial $\chi_i$ and final $\chi_f$ states as a function of relative cluster motion with the intercluster distance *R*.

Using the formula for the magnetic transition *M*1(*S*) caused by the spin part of the magnetic operator we can obtain ($S_i = S_f = S$, $L_i = L_f = L$) [7]:

$$P_1^2(M1, J_f, J_i) = \delta_{S_i S_f} \delta_{L_i L_f} \left[ S(S+1)(2S+1)(2J_i+1)(2J_f+1) \right] \begin{Bmatrix} S & L & J_i \\ J_f & 1 & S \end{Bmatrix}^2 ,$$

$$A_1(M1, K) = i \frac{\hbar K}{m_0 c} \sqrt{3} \left[ \mu_1 \frac{m_2}{m} - \mu_2 \frac{m_1}{m} \right], \quad I_J(J_f, J_i) = \langle \chi_f | R^{J-1} | \chi_i \rangle , \quad J = 1. \quad (25)$$

Here $m$ is the mass of the nucleus; $\mu_1$ and $\mu_2$ are the magnetic moments of the clusters, the rest notations – as in the previous expression. The values: $\mu_p = 2.792847356 \mu_0$, $\mu_n = -1.91304272 \mu_0$, $\mu(^2H) = 0.857438231 \mu_0$ [34] are used for magnetic moments of proton, neutron, and deuteron. The correctness of the given above expression for the $M1$ transition preliminarily checked on the basis of the radiative proton capture reactions on $^2H$ and $^7Li$ at low energies in our works [7,15].

The principle of detailed balance is used for finding of the photodisintegration cross section [8]

$$\sigma_d(J_0) = \frac{q^2}{K^2} \frac{(2S_1+1)(2S_2+1)}{2(2J_0+1)} \sigma_c(J_0) , \quad (26)$$

where $J_0$ – the total moment of nucleus, $\sigma_c$ – the total radiative capture cross section, $\sigma_d$ – the photodisintegration cross sections.

Exact mass values of particles generally equal to: $m_p = 1.00727646677$, $m_n = 1.00866491597$, $m_{2H} = 2.013553212724$ amu were specified in the calculation of the radiative capture [34].

## 2.5 Construction of the intercluster potentials

Dwell more thoroughly on the procedure of construction of the intercluster partial potentials used here of the form (2) or (3) at the given orbital moment $L$, estimating criteria and sequence of parameter finding and pointing to their errors and ambiguities.

The parameters of the BS potentials are found in the first place, if they at the given number of the allowed and forbidden states in this partial wave are fixed quite unambiguously according to binding energy, nuclear radius and asymptotic constant in the considered channel. The accuracy of the obtained BS parameters is connected, in the first place, with the AC accuracy, which is usually equal to 10÷20%, because the accuracy of the experimental determination of the charged radius usually much higher – 3÷5%. There are no other ambiguities in this potential, because the classification of states according to Young tableaux allows one unambiguously to fix the number of BSs, FSs, or ASs in this partial wave, which completely determine its depth, and the potential width depends wholly from the AC value. The principles of determination the number of FSs and ASs for given partial wave are cited in the next paragraph. It is necessary to note here that calculations of the charged radius in any model have model errors, i.e., errors that caused by the accuracy of the model itself. In the MPCM the value of this radius depends from the integral over the model WFs, i.e., the model errors of such functions simply sum. The AC values are determined by the asymptotics of model WFs in one point at their asymptotics and, evidently, have appreciably lower error. Therefore, in future the BS potentials are constructed so that, in the first place, maximally agree with the AC values that were obtained on the basis of independent methods, which allow one to extract AC from the experimental data [24].

The intercluster potential of the nonresonance scattering process obtained according to the scattering phase shifts at the given number of BSs, ASs, and FSs in the considered partial wave also constructs quite unambiguously. The accuracy of the determination of this potential connects with, in the first place, the accuracy of phase shift extraction scattering phases from experimental data and can reach 20÷30%. And here this potential does not have ambiguities, because the classification of the states according to Young tableaux allows one unambiguously to fix the number of the BSs, which absolutely determines its depth, and the width of the potential at the given depth is determined by the shape of the scattering phase shift. At the construction of the nonresonance scattering potential according to data on nuclear spectra in the certain channel, it is difficult to estimate the accuracy of determination of its parameters even at the given number of BSs, though it is possible, evidently, to hope that it a little higher the error in the previous case. This potential, as it is usually supposed for the energy range down to 1 MeV, has to lead to the scattering phase shift approximate to zero or gives the taper shape of the phase shift, because there are no resonance levels in the spectrum of nucleus.

At the analysis of the resonance scattering, when in the considered partial wave at energies down to 1 MeV there is a rather sharp resonance with the width of 10÷50 keV, at the given number of BSs (FSs or ASs), the potential also constructs completely unambiguously. At the given number of BSs its depth fixes unambiguously according to the resonance energy of the level, and the width is completely determined by the width of such resonance. The error of its parameters usually not exceed the error of width determination of this level and equals approximately 3÷5%. Meanwhile this applies to the construction of the partial potential according to the scattering phases and to the determination of its parameters by the resonance in nuclear spectra.

Consequently, all potentials have not ambiguities and allow one to describe total cross sections of the radiative capture processes, without involvement of such notation as the spectroscopic factor $S_f$, i.e., its value simply to take equal to unit, as it was done in work [33]. In other words, at the consideration of the capture reaction in the MPCM for the potentials matched in the continuous spectrum with the characteristics of scattering process which take into account the resonance shape of the phase shifts and with the characteristics of discrete spectrum describing the basic properties of the BS of nucleus, so there is no necessity to introduce additional term $S_f$ [33]. Evidently, all available effects in this reaction, including the possibility of the cluster configuration are taking into account at the construction of the interaction potentials. It becomes possible because the potentials are constructed taking into account the structure of the FSs and on the basis of description observable, i.e., experimental characteristics of interacting clusters in the initial channel and formed in the final state some nucleus, which describing by the cluster structure that consists with initial particles. Thus, the presence of the $S_f$, takes into account into the BS cluster wave functions, which become clear on the basis of such potentials by solving the Schrödinger equation (6).

In conclusion we note that at the construction of partial interaction potentials it is taken into consideration that they depend not only from orbital moment $L$, but from total spin $S$ and total moment $J$ of the cluster system. In other words, we will have different parameter values for different moments $L$, $S$, $J$. Since, usually the $E$1 and the $M$1 transitions between different states $^{(2S+1)}L_J$ in continuous and discrete spectra are considered, so the potentials of these states will be different.

## 2.6 Classification of the cluster states

States with minimal spin in the processes of scattering of some light atomic nuclei turn out to be mixed according to Young orbital tableaux, for example, the doublet state p$^2$H [6] is mixed according to tableaux {3} and {21}. On the other hand, these states considered as bound ones, for example, the doublet p$^2$H channel of $^3$He nucleus is pure with tableau {3} [6]. Let us put the classification of states of, for example, p$^2$H system according to orbital and spin-isospin Young tableaux and demonstrate how to obtain these results. In the general case, the possible orbital Young tableau {$f$} of some nucleus $A(\{f\})$ consisting of two parts $A_1(\{f_1\}) + A_2(\{f_2\})$ is the direct outer product of orbital Young tableaux of these parts $\{f\}_L = \{f_1\}_L \times \{f_2\}_L$ and is determined using the Littlewood's theorem [6,22]. Therefore, the possible orbital Young tableaux of p$^2$H system, in which tableau {2} is used for $^2$H nucleus, are the symmetries {3}$_L$ and {21}$_L$.

Spin-isospin tableaux are the direct inner product of spin and isospin Young tableaux of the nucleus from $A$ nucleons $\{f\}_{ST} = \{f\}_S \otimes \{f\}_T$ and for the system with the number of particles not larger than eight are given in [35]. For any of these moments (spin or isospin), the corresponding tableau of the nucleus consisting of $A$ nucleons, each of which has an angular moment equals 1/2, is constructed as follows: in the cells of the first row, the number of nucleons with the moments pointing in one direction, for example, upward, is indicated. In cells of the second row, if it is required, the number of nucleons with the moments directed in the opposite direction, for example, downward, is indicated. The total number of cells in both rows is equal to the number of nucleons in the nucleus. Moments of nucleons in the first row which have a pair in the second row with the oppositely directed moment are compensated and yield zero total moment. The sum of nucleon moments of the first row, which are not compensated by moments of nucleons of the second row, yields the total moment of the whole system. In this case for simplest $N^2$H cluster system at the isospin of $T = 1/2$ we have {21}$_T$; for the spin state of $S = 1/2$, we also obtain {21}$_S$; and for $S$ or $T = 3/2$, the Young tableau have the form {3}$_{ST}$. Upon construction of the spin-isospin Young tableau for the quartet spin state of $N^2$H system with $T = 1/2$, we have {3}$_S \otimes$ {21}$_T$ = {21}$_{ST}$, and for the doublet state {21}$_S \otimes$ {21}$_T$ = {111}$_{ST}$ + {21}$_{ST}$ + {3}$_{ST}$ [35].

The total Young tableau of the nucleus is determined in a similar way as the direct inner product of the orbital and spin-isospin tableau $\{f\} = \{f\}_L \otimes \{f\}_{ST}$ [22]. The total wave function of the system in the case of antisymmetrization does not identically vanish only if it does not contain the antisymmetric component $\{1^N\}$, which is realized upon multiplication of conjugated $\{f\}_L$ and $\{f\}_{ST}$. Therefore, the tableaux $\{f\}_L$ conjugated to $\{f\}_{ST}$ are allowed in this channel and all other symmetries are forbidden, since they result in zero total wave function of the system of particles after its antisymmetrization. This yields that, for the $N^2$H systems in the quartet channel, only the orbital wave function with the symmetry {21}$_L$ is allowed and the function with {3}$_L$ turns out to be forbidden, since the product {21}$_{ST} \otimes$ {3}$_L$ does not result in an antisymmetric component of the wave function. At the same time, in the doublet channel, we have $\{111\}_{ST} \otimes \{3\}_L = \{111\}$ and $\{21\}_{ST} \otimes \{21\}_L \sim \{111\}$ [35], and in both cases we obtain the antisymmetric tableau. Therefore, the doublet spin state turns out mixed according to Young orbital tableaux.

In [6,22] the method for separation of such states according to Young tableaux was proposed and it was shown that the mixed scattering phase shifts can be represented in the form of the half-sum of pure phase shifts $\{f_1\}$ and $\{f_2\}$ [6],

$$\delta^{\{f_1\}+\{f_2\}} = 1/2(\delta^{\{f_1\}} + \delta^{\{f_2\}}) \qquad (27)$$

In this case it is assumed that $\{f_1\} = \{21\}$ and $\{f_2\} = \{3\}$ and the doublet phase shifts extracted from the experiment are mixed according to these two tableaux. Then it is assumed that the quartet scattering phase shifts, pure according to orbital Young tableau $\{21\}$, can be identified with the pure doublet scattering phase shift $N^2H$ corresponding to the same Young tableau. Then Eq. (27) allows to find the pure doublet $N^2H$ phase shift with tableau $\{3\}$ and then construct the pure interaction potential according to Young tableaux that can be used for description of characteristics of the bound state.

Really, the potential of the doublet BS is constructed so that to correctly reproduce the basic characteristics of this state. Meanwhile, it is considered that it corresponds to the pure state with one Young tableau. The doublet scattering states are considered as mixed according Young tableaux and are constructed based on the correct description of the correspondent scattering phase shifts. In this case the potentials of the BS and scattering for the doublet states were different from the difference of Young tableaux for such states. In other words, the obvious dependence of the potential parameters at the given $L$, $S$ and $J$ from Young tableaux $\{f\}$ is assumed.

## 3 Astrophysical *S*-factor of the radiative proton capture on $^2$H

Let us start the consideration of thermonuclear reactions from the radiative capture process

$$p + {}^2H \rightarrow {}^3He + \gamma, \qquad (28)$$

which is the first nuclear reaction of the proton-proton or pp-chain following due to the electromagnetic interactions, since γ-quantum takes a part in it [36]. This reaction process makes essential contribution into the energy yield of the fusion reactions [37] that, as usually considered, determine the burning of the Sun and stars of our Universe. Since, interacting nuclear particles of the proton-proton chain have a minimal potential barrier. The pp-chain is the first chain of nuclear reactions which can take place at lowest energies and, consequently, at stellar temperatures, and there is in all stable stars of the Main Sequence.

The radiative capture process on $^2$H in the pp-chain is the basic one for the transition from the primary proton fusion

$$p + p \rightarrow {}^2H + e^+ + \nu_e, \qquad (29)$$

which takes place due to the weak interactions with the participation of electron neutrino $\nu_e$ to one of the final reaction of capture in pp-chain of two nuclei $^3$He [38]

$$^3\text{He} + {}^3\text{He} \to {}^4\text{He} + 2p, \qquad (30)$$

which occurs due to the strong nuclear interactions [36].

A detailed study of the radiative proton capture on $^2$H reaction from the theoretical and experimental points of view has the fundamental interest not only for nuclear astrophysics, but for the whole nuclear physics of ultralow energies and the lightest atomic nuclei [15]. Therefore, experimental studies of this process are continuing and already at the beginning of $2000^{\text{th}}$ year, due to the European project LUNA, new experimental data of the radiative proton capture on $^2$H at energies down to 2.5 keV has appeared. These energies may take place in fusion reactions in the Sun and many stable stars [38]. These experimental results, along with earlier results at greater energies will be used by us furthermore for comparison with the results of our calculations.

It should be noted that the lightest nuclei with A≤4, strictly speaking, are neither shell nor cluster ones. This follows from microscopic calculations of these nuclei with realistic $NN$ potentials (see, for example, [39]). For example, in an $^3$He nucleus along with the p$^2$H cluster configuration, the configuration p$^2$H* is also present, where $^2$H* is the spin-singlet deuteron (np pair in $^1S_0$ state), and spectroscopic factors for common and singlet deuterons are approximately equal to $S = 1.5$ [40,41]. The channel with singlet deuteron is clearly manifested in the elastic p$^3$He backward scattering both in purely nucleon scattering mechanism [40] and in processes with production of virtual π meson [41]. However, at low energies and low momentum transfers, it is reasonable to apply the considered two-cluster approach to few-nucleon systems with $A = 3$ and 4, at least to compare results obtained in the frame of the MPCM with multi-body calculations (see, for example, [42,43]) and the MPCM results for systems with A>4. In this regard, the using of this approach to such systems, especially for low-energy process analysis, seems quite reasonable.

## 3.1 Potentials and scattering phase shifts

The total cross sections of photoprocesses for lightest $^3$He and $^3$H nuclei in the potential cluster model with FSs were considered earlier in our works [12]. In these calculations for photodisintegration of $^3$He and $^3$H into the p$^2$H and n$^2$H channels, the $E1$ transitions caused by the orbital part of the electric operator $Q_{Jm}(L)$ [8] were taken into account. Cross sections of the $E2$ processes and cross sections depending on the spin part of electric operator turned out to be lower by several orders of magnitude. Then it was assumed that electric $E1$ transitions in the $N^2$H system are possible between the ground doublet $^2S$ state pure according to Young tableau {3} of $^3$H and $^3$He nuclei and doublet $^2P$ scattering states mixed according to Young tableaux {3} + {21}. Such transition is quite possible since the quantum number, connected with Young tableaux, evidently is not saved in electromagnetic processes [6,22].

Furthermore, for calculation of photonuclear processes in p$^2$H and n$^2$H systems [21], the nuclear part of the intercluster interaction potential was represented in form (2) with the point-like Coulomb term, the Gaussian attractive $V_0$, and the exponential repulsive $V_1$ parts. The potential of each partial wave was constructed in order to correctly describe the corresponding partial elastic scattering phase [44]. Using these representations the potentials of the p$^2$H interaction for scattering processes were obtained; the parameters of these potentials are given in [8,12] and second and third rows in Table 1.

Table 1. Potentials of the p$^2$H interaction [12] for spin of $S = 1/2$.

| $^{2S+1}L, \{f\}$ | $V_0$, MeV | $\alpha$, fm$^{-2}$ | $V_1$, MeV | $\gamma$, fm$^{-1}$ |
|---|---|---|---|---|
| $^2S, \{3\}$ | -34.76170133 | 0.15 | – | – |
| $^2S, \{3\}+\{21\}$ | -55.0 | 0.2 | – | – |
| $^2P, \{3\}+\{21\}$ | -10.0 | 0.16 | +0.6 | 0.1 |

With kind permission of the European Physical Journal (EPJ).

The caused out calculations of the $E1$ transition showed [12] that that it is possible to describe the total photodisintegration cross sections for $^3$He nucleus in the region of γ-quantum energies 6–28 MeV, including the maximum at $E_\gamma$ = 10–13 MeV. The potential of the $^2P$ wave of the p$^2$H scattering with peripheral repulsion given in Table 2.1 and the $^2S$ interaction of the bound state, pure according to Young tableau {3}, which has a Gaussian form with zero repulsion $V_1 = 0$ and with the parameters $V_0 = -34.75$ MeV, $\alpha = 0.15$ fm$^{-2}$ of the attractive part are used for these calculations. These parameters were obtained based on a correct description of the binding energy (with an accuracy of several keV) and charge radius of $^3$He.

The calculations of the total cross sections of the proton radiative capture on $^2$H and astrophysical $S$-factor at energies from 10 keV to 1.0 MeV [8,12] were carried out with such potentials. These results acceptably describe the available at that moment experimental data on the $S$-factor at the energy lower than 1.0 MeV [45]. However, in this work there were carried out measurements only at the energy down to 170 keV. Thus, at that moment we knew the experimental measurements on the $S$-factor of the proton radiative capture on $^2$H only in the energy range above 170 keV [45]. Comparatively recently the new experimental data on the $S$-factor of the proton radiative capture on $^2$H at the energy from 2.5 keV to 170 keV [46-48] were appeared. It was found after their analysis that the calculations which were done earlier are based on the $E1$ process only, which coincide with them completely in the considered energy range from 10–20 keV to 1.0 MeV [15]. Thereby, the using potential cluster model is able to describe new experimental data, but also, per se, predict the behavior of the astrophysical $S$-factor of the proton capture on $^2$H beforehand at the energy range down to 10–20 keV. The calculations presented in 1995 year in our works [12] were done before carrying out new experimental measurements [48] in 2002 year and even the results of more early works [46,47], published in 1997.

Furthermore, the parameters of the "pure" doublet $^2S$-potential according to Young tableau {3} [15] were adjusted for a more accurate description of the experimental binding energy of $^3$He in the p$^2$H channel. This potential (see Table 2.1) has become somewhat deeper than the potential we used in our works [12] and leads to the total agreement between calculated -5.4934230 MeV and experimental -5.4934230 MeV [49] binding energies obtained with exact values of particle masses [34]. The difference between the potentials given in [12] and in Table 1 [15] is primarily due to the application of the exact masses of particles and more accurate description of binding energy of $^3$He in the p$^2$H channel. This difference of parameters for depths of potentials is equal to 0.012 MeV and has not any influence on the calculation results for S-factor at any considered energies.

The charge radius of $^3$He with this potential equals 2.28 fm, which is a little higher than the experimental values listed in Table 2 [34,49,50]. As one can see from this data, the radius of the deuteron cluster is larger than radius of $^3$He. Thus, if the deuteron is located in $^3$He as a cluster, it must be compressed by about 20-30% of its own size in a free state for a correct description of the $^3$He charge radius [8,29,51].

Table 2. Experimental masses and charge radii of light nuclei used in these calculations [34,49,50].

| Nucleus | Radius (fm) | Mass (amu) |
|---|---|---|
| $^1$H | 0.8768(69) | 1.00727646677 |
| $^2$H | 2.1402(28) | 2.013553212724 |
| $^3$H | 1.63(3); 1.76(4); 1.81(5) <br> The average value is 1.73 | 3.0155007134 |
| $^3$He | 1.976(15); 1.93(3); 1.877(19); 1.935(30) <br> The average value equals 1.93 | 3.0149322473 |
| $^4$He | 1.671(14) | 4.001506179127 |

In order to control the behavior of wave functions of bound states at large distances, the asymptotic constant $C_W$ with the asymptotics of the wave function in the form of Whittaker function (5) was calculated; the value of this constant in an interval 5–20 fm is equal to $C_W = 2.33(3)$. The error shown here is determined by averaging the constant over the interval indicated above.

The determination of this constant from the experimental data yields values in an interval 1.76–1.97 [52-54] that is somewhat lower than the value obtained here. The results of three-body calculations [55] should also be mentioned; in these calculations, good agreement with experiment [56] for the ratio of asymptotic constants of $^2S$ and $^2D$ waves was obtained and the following constant value of $C_W = 1.878$ was obtained for the $^2S$ wave oneself. However, in work [23] that was published later than [52-54], a value 2.26(9) was given for $C_W$, which agrees well with our calculations. It can be seen from the data presented in these works that experimental results on asymptotic constants obtained at different times and by different authors scatter considerably. These data are in the range from 1.76 to 2.35 with an average value 2.06. In one of the last work [57] devoted to the extraction of the $C_W$ constants from the experimental data (with the given in this work refs. on other results) were obtained 2.25(13) with $\sqrt{2k_0} = 0.918$ that within the limits of error almost coincide with the given above average AC value and absolutely agree with the results of work [23]. Slightly different definition of AC: $\chi_L(r) = C_0 W_{-\eta L+1/2}(2k_0 r)$ was used in work [57], it differs from our (5) by the term $\sqrt{2k_0}$ for which the value 2.07(12) fm$^{-1/2}$ was obtained – in this case $C_W = C_0/\sqrt{2k_0}$.

In the potential two-cluster model, the $C_W$ value and the charge radius strongly depend on the width of the potential well. Other parameters of the ground state $^2S$ potential can always be found, for example,

$$V_0 = -48.04680730 \text{ MeV and } \alpha = 0.25 \text{ fm}^{-2}, \tag{31}$$

$$V_0 = -41.55562462 \text{ MeV and } \alpha = 0.20 \text{ fm}^{-2}, \tag{32}$$

$$V_0 = -31.20426327 \text{ MeV and } \alpha = 0.125 \text{ fm}^{-2}, \tag{33}$$

which yield the same binding energy for $^3$He in the p$^2$H channel. The first of them at an interval 5–20 fm results in the asymptotic constant $C_W = 1.945(3)$ and the charge radius $R_{ch} = 2.18$ fm, the second yields the constant $C_W = 2.095(5)$ and $R_{ch} = 2.22$ fm, and the third $C_W = 2.519(3)$ and $R_{ch} = 2.33$ fm. The cluster radii from Table 3.2 are used in the calculations of charge radii.

It can be seen from these results that potential (31) makes it possible to obtain the charge radius that is the closest to experiment value. Further reduction of the potential width may result in correct description of its value; however, it will be shown below that it will not make it possible to reproduce the S-factor of radiative p$^2$H capture. In this sense potential (32), which is characterized by somewhat larger width, has the minimal admissible width of the potential well for which it is possible to obtain an asymptotic constant practically equal to its experimental average value 2.06 and acceptably describe the behavior of the astrophysical S-factor in the broadest energy range.

For a complementary check of the determination of binding energy in two-body channels the variational method (VM) with the expansion of WF on the nonorthogonal Gaussian basis (10) and with independent parameter variation [7] has been used. This method already made it possible to obtain a binding energy of –5.4934228 MeV for a grid with a dimensionality 10 for the pure, according Young tableaux, potential from Table 1. The asymptotic constant $C_W$ of the variational wave function at distances 5–20 fm in these calculations was on a level 2.34(1), and the residual did not exceed $10^{-12}$ [7]. The parameters and coefficients of expansion of the radial wave function for this potential of form (27) are given in Table 3. Under the finding of the coefficients of expansion of the WF $C_i$ they are determined so that to lead to the normalization of the WF equals 1 [7]. The normalization coefficients $N$, given in this table and for all similar results, determine, per se, the accuracy of finding of such coefficients in the VM, i.e., the accuracy of normalization of the WF to unit.

Table 3. Variational parameters and expansion coefficients of radial wave function of the p$^2$H bound state system for potential from Table 1. Normalization of function with these coefficients on an interval 0–25 fm is $N = 0.999999997$

| $i$ | $\beta_i$ | $C_i$ |
| --- | --- | --- |
| 1 | 2.682914012452794E-001 | -1.139939646617903E-001 |
| 2 | 1.506898472480031E-002 | -3.928173077162038E-003 |
| 3 | 8.150892061325998E-003 | -2.596386495718163E-004 |
| 4 | 4.699184204753572E-002 | -5.359449556198755E-002 |
| 5 | 2.664477374725231E-002 | -1.863994304088623E-002 |
| 6 | 4.468761998654231E+001 | 1.098799639286601E-003 |
| 7 | 8.482112461789261E-002 | -1.172712856304303E-001 |
| 8 | 1.541789664414691E-001 | -1.925839668633162E-001 |
| 9 | 1.527248552219977E-000 | 3.969648696293301E-003 |
| 10 | 6.691341326208045E-000 | 2.097266548250023E-003 |

Table 4. Variational parameters and expansion coefficients of radial wave function of the p²H bound state system for potential (32). Normalization of function with these coefficients on an interval 0–25 fm is $N = 0.999999998$

| i | $\beta_i$ | $C_i$ |
|---|---|---|
| 1 | 3.485070088054969E-001 | -1.178894628072507E-001 |
| 2 | 1.739943603152822E-002 | -6.168137382276252E-003 |
| 3 | 8.973931554450264E-003 | -4.319325351926516E-004 |
| 4 | 5.977571392609325E-002 | -7.078243409099880E-002 |
| 5 | 1.245586616581442E-002 | -2.743665993408441E-002 |
| 6 | 5.837991732045449E+001 | 1.102401456221556E-003 |
| 7 | 1.100441373510820E-001 | -1.384847981550261E-001 |
| 8 | 2.005318455817479E-001 | -2.114723533577409E-001 |
| 9 | 1.995655373133832E-000 | 3.955231655325594E-003 |
| 10 | 8.741651544040529E-000 | 2.101576342365150E-003 |

With kind permission of the European Physical Journal (EPJ).

The variant of potential (32) was also considered in the framework of the variational method; the same binding energy, –5.4934228 MeV, has been obtained for this potential. The variational parameters and coefficients of expansion of the radial wave function are given in Table 4. The asymptotic constant in a range 5–20 fm turned out to be equal to 2.09(1), and the residual was of the order of $10^{-13}$.

Since the variational energy decreases with increasing dimensionality of the basis and yields the upper boundary of the true binding energy [58], and the finite-difference energy increases with decreasing step and increasing number of steps [7], an average value of –5.4934229(1) MeV can be taken as a realistic estimate of the binding energy in this potential. Thus, it may be considered that the error of determination of the binding energy of the p²H system in ³He nucleus using two methods based on two different computer programs is ±0.1 eV in the given potential.

### 3.2 Astrophysical S-factor

In our calculations of the astrophysical S-factor [15] of the radiative proton capture on ²H the region of energies from 1 keV to 10 MeV was considered and also the E1 transition from the ²P wave of scattering to the ground ²S state with {3} and potential parameters listed in Table 1. For the S(E1)-factor for 1 keV, a value of 0.165 eV b was obtained, which is in a quite agreement with the known data including the separation of S(0)-factor into $S_s$ and $S_p$ parts due to M1 and E1 transitions. This separation was made in [47], where it was obtained that $S_s(0) = 0.109(10)$ eV b and $S_p(0) = 0.073(7)$ eV b, which for the total S-factor should yield 0.182(17) eV b. On the other hand, in the expression for linear interpolation of the total S-factor,

$$S(E_{c.m.}) = S_0 + E_{c.m.}S_1, \qquad (34)$$

the authors give the following values [47]: $S_0 = 0.166(5)$ eV b and $S_1 = 0.0071(4)$ eV b keV$^{-1}$, and for S(0) a value 0.166(14) keV b was given; this value is determined with all possible errors taken into account. The results obtained with separation of the *S*-factor into *M*1 and *E*1 parts were given in one of the first papers [45] devoted to astrophysical factors, wherein it was obtained that $S_s(0) = 0.12(3)$ eV b and $S_p(0) = 0.127(13)$ eV b for a total *S*-factor 0.25(4) eV b. These data of $S_s(0)$ value, within the limits of errors, quite coordinate with the data given in [47].

Experimental data one of the last work [48] yield the total astrophysical factor of $S(0) = 0.216(10)$ eV b; this means that the contributions of *M*1 and *E*1 differ from the above values [47]. In this paper the following parameters of linear extrapolation (34) are given: $S_0 = 0.216(6)$ eV b and $S_1 = 0.0059(4)$ eV b keV$^{-1}$, which noticeably differ from the data of [47]. The other known results for the *S*-factor obtained from experimental data without separation into *M*1 and *E*1 parts yield for zero energy 0.165(14) eV b [59]. Previous results of the same authors yield 0.121(12) eV b [60], and in theoretical calculations [61] the following values were obtained for different models: $S_s(0) = 0.105$ eV b and $S_p(0) = 0.08$–0.0865 eV b and the total *S*-factor equals to 0.185–0.192 eV b. It follows from these results that there exists a great ambiguity in the data obtained during the last 20 years. These results make it possible to conclude that, most probably, the value of the total *S*-factor at zero energy is in an interval from 0.109 eV b [60] to 0.226 eV b [48]. The average of these values yields an *S*-factor equal to 0.167(59) eV b, which quite agrees with that obtained here based on *E*1 transition.

Our calculations of the *S*(*E*1)-factor of the radiative proton capture on $^2$H for the potential given in Table 1 at energies from 1 keV to 10 MeV are shown in Figs. 1 and 2 with dotted lines. The obtained *S*-factor rather well reproduces new experimental data at energies 10–50 keV [47], and at lower energies the calculated curve is within the interval of experimental errors of [48].

The solid lines in Figs. 1 and 2 show the results for potential (32), which reproduces the behavior of *S*-factor at energies 50 keV–10 MeV somewhat better and for 1 keV yields $S_p = 0.135$ eV b. For 20–50 keV, the calculated curve follows the lower boundary of errors [47], and below 10 keV it falls into the interval of experimental errors of the project LUNA obtained most recently [48]. The value of the *S*-factor obtained at zero energy with this potential agrees well with data [45] for the electric *E*1 transition $S_p(0)$.

The dashed lines in Figs. 1 and 2 show the results for potential (33), and the dash-dotted line, for potential (31). It can be assumed based on these calculations that the best results are obtained for bound state potential (32), which describes experimental data in the broadest energy interval. It provides a certain compromise in description of the asymptotic constant, charge radius, and astrophysical *S*-factor of radiative p$^2$H capture.

It can be seen in Fig. 1 that, at low energies of about 1–3 keV, the *S*-factor for potential (32) is practically independent of the energy; thus, the determination of this factor at zero energy yields approximately the same value as at 1 keV. Therefore, the difference of *S*-factor at 0 and 1 keV probably comes to no more than 0.005 eV b; this quantity can be assumed for the error of determination of the calculated *S*-factor for zero energy and accept that it equals 0.135(5) eV b.

The dashed lines in Figs. 1 and 2 show the results for potential (33), and the dash-dotted line, for potential (31). It can be assumed based on these calculations that the best results are obtained for bound state potential (32), which describes experimental data in the broadest energy interval. It provides a certain compromise in description of the asymptotic constant, charge radius, and astrophysical $S$-factor of radiative $p^2H$ capture.

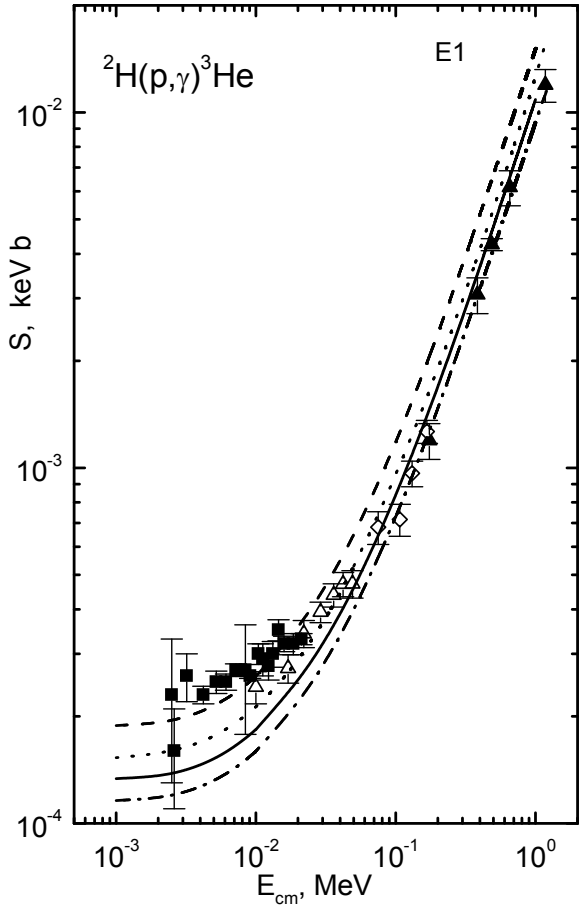 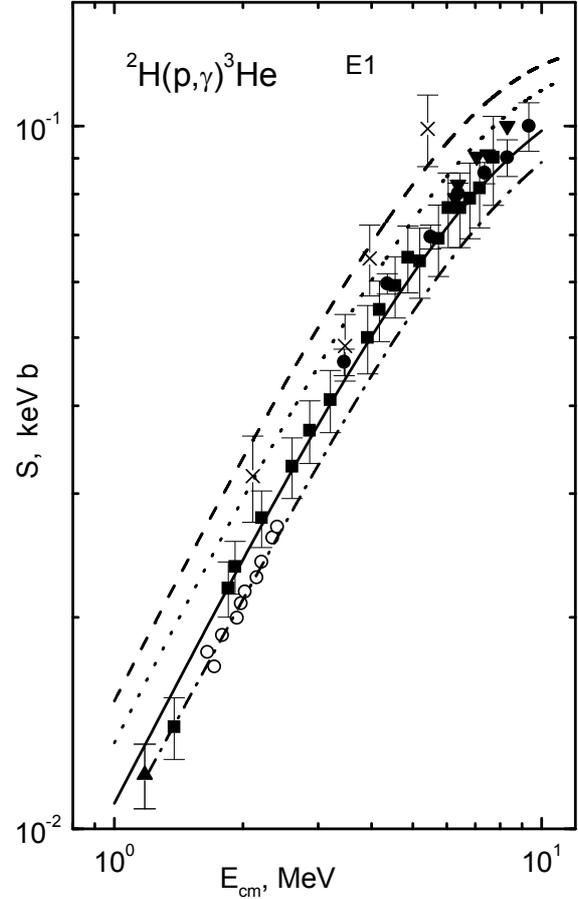

Fig. 1. Astrophysical $S$-factor of the radiative proton capture on $^2$H in a range 1 keV–1 MeV for the $E$1 transition. Curves show calculations with potentials given in the text. Triangles show experiment [45], open rhombs [46], open triangles [47], and squares [48].

Fig. 2. Astrophysical $S$-factor of the radiative proton capture on $^2$H in a range 1–10 MeV for the $E$1 transition. Curves show calculations with potentials given in the text. Upward triangles show experiment from [45], squares [62], points [63], crosses [64], downward triangles [65], and circles [66].

With kind permission of the European Physical Journal (EPJ).

It can be seen in Fig. 1 that, at low energies of about 1–3 keV, the $S$-factor for potential (32) is practically independent of the energy; thus, the determination of this factor at zero energy yields approximately the same value as at 1 keV. Therefore, the difference of $S$-factor at 0 and 1 keV probably comes to no more than 0.005 eV b; this quantity can be assumed for the error of determination of the calculated $S$-factor for zero energy and accept that it equals 0.135(5) eV b.

At low energies, the $M$1 transition from the $^2S$ scattering state mixed according to Young tableaux to the bound $^2S$ state of $^3$He pure for orbital symmetry can give contribution to the total astrophysical $S$-factor. For these calculations we used doublet $^2S$ potential of scattering states with parameters given in Table 1 [12,15] and the $^2S$ GS potential with parameters (32). We would remind you that due to the different Young tableaux for these states their potentials can be different.

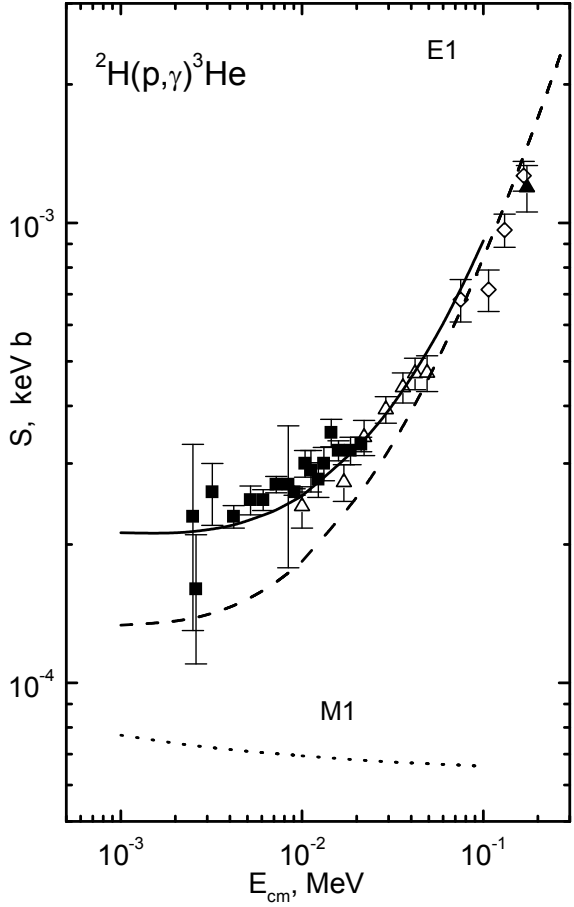

Fig. 3. Astrophysical *S*-factor of the radiative proton capture on $^2$H in a range 1 keV–0.3 MeV for $E$1 and $M$1 transitions. Curves show calculations with potentials given in the text. Triangles show experiment from [45], open rhombs [46], open triangles [47], and open squares [48].

The results of calculations of the $M$1 process at 1–100 keV are shown in Fig. 3 by the dotted line at the bottom of the figure and the results of $E$1 transition for the GS potential with the parameters (32) are shown by the dashed line – they are shown in Fig. 1 by the solid line. The total *S*-factor is shown in Fig. 3 by the solid line, which well demonstrates a small contribution of $M$1 into *S*-factor at the energies above 100 keV and its significant influence to the energy range of the order of 1–50 keV.

The energy dependence of the total *S*-factor in the range of 2.5–50 keV is in complete accordance with the findings of works [47,48] and for the *S*-factor of the $M$1 transition at 1 keV we obtained the value 0.077 eV b, which leads to the value 0.212(5) eV b for the total *S*-factor and which is in a good agreement with the new measurements data from LUNA project [48]. And as it can be seen from Fig. 3, at the energies 1–3 keV the value of the total *S*-factor is more stable than it was for the $E$1 transition and we consider it to be absolutely reasonable to write the result as 0.212 eV b with an error 0.005 eV b.

If expression (34) will be used for the *S*-factor parametrization, then it is possible to describe the solid line in Fig. 3 by the parameters $S_0 = 0.1909$ eV b and $S_1 = 0.006912$ eV b keV$^{-1}$ in the energy range of 1–100 keV, with the average $\chi^2 = 0.055$. If we use the quadric form of parametrization

$$S(E_{c.m.}) = S_0 + E_{c.m.}S_1 + E^2_{c.m.}S_2, \qquad (35)$$

The next values were obtained for the parameters: $S_0=0.1957$ eV b, $S_1=0.006055$ eV b keV$^{-1}$ and $S_2=0.00001179$ eV b keV$^{-2}$, with the average $\chi^2=0.017$ in the energy range 1-100 keV. The 10% errors of the calculated *S*-factor values are used for determination of $\chi^2$. The approximation of the calculation results, by the analytical function of certain type with the performing of $\chi^2$ minimization, is actually done here and further, therefore the $S_0$ and $S(0)$ values are slightly differ, but this difference usually not more than 10%. The quadratic form (35) reproduces the behavior of the calculated *S*-factor a bit better, as one can see. There is another method of the *S*-factor parametrization: when the value $S_0$, determining its behavior at zero energy, is predetermined. In this case, these values are obtained for the parameters of the form (35): $S_0=0.2120$ eV b, $S_1=4.5366 \cdot 10^{-3}$ eV b keV$^{-1}$ and $S_2=2.8622 \cdot 10^{-5}$ eV b keV$^{-2}$ with the average $\chi^2=0.124$, at the same energy range and for 10% errors.

Furthermore, it is necessary to note that we are unable to build the scattering $^2S$-potential uniquely, because of the ambiguities in the results of phase shift analysis of the p$^2$H scattering. The other variant of the potential with parameters $V_0$=-35.0 MeV and $\alpha$=0.1 fm$^{-2}$ [8,12], which also describes well the $S$ phase shift of scattering, leads at these energies to $S$-factor of the $M$1 process several times lower than in the previous case. Such a big ambiguity in parameters of the $^2S$-potential of scattering, associated with errors of phase shifts extracted from the experimental data, does not allow us to make certain conclusions about the contribution of the $M$1 process in the radiative proton capture on $^2$H, although the first of the described calculation variants are matched with the latest measurements [47,48].

If the GS potentials are determined by the binding energy, asymptotic constant and charge radius, and also by an additional criterion – use of interactions "pure" in accordance with Young tableaux quite uniquely and the potential description of the scattering phase shifts, which are "pure" in accordance with Young tableaux, but the situation with the construction of scattering potentials is not so unambiguous. Then, in the case of scattering, it is necessary to carry out a more accurate phase shift analysis for the $^2S$ wave and to take into account the spin-orbital splitting of the $^2P$ phase shifts at low energies, as it was done for the elastic p$^{12}$C scattering at energies 0.2–1.2 MeV [67]. Carrying out of this additional analysis will allow us to adjust the potential parameters used in the calculations of the p$^2$H capture in the potential cluster model, thereby increasing the accuracy of the calculation results.

Thus, the $S$-factor calculations of the proton radiative capture on $^2$H for the $E$1 transition at the energy range down to 10 keV, which we carried out about 20 years ago [12], when the experimental data above 150–200 keV [45] was only known, are in a good agreement with the new data of works [46,47] in the energy range from 10–20 to 150–200 keV. Meanwhile, this concerns the GS potential from Table 1, which, per se, was used in works [12], and the interaction with parameters from (32). The results of $S_p$-factors at the $E$1 transitions for two considered potentials at the energies lower than 10 keV (see Fig. 1) practically fall within the error band of work [48] and show that the $S$-factor tends to remain constant at energies 1–3 keV.

In spite of the uncertainty of the $M$1 contribution to the process, which results from the errors and ambiguity of $^2S$ scattering phases, the scattering potential from Table 1 with mixed Young tableaux in the $^2S$ wave allows one to obtain a reasonable value for the astrophysical $S_s$-factor of the magnetic transition in the range of low energies. At the same time, the value of the total $S$-factor is in a good agreement with all known experimental measurements [46-48] at energies from 2.5 keV to 10 MeV (Figs. 1 and 3).

As a result, the MPCM based on the intercluster potentials adjusted for the elastic scattering phase shifts and GS characteristics, for which the FS structure is determined on the classification of BSs according to Young orbital tableaux and with the parameters suggested as early as 20 years ago [12], allows one correctly to describe the astrophysical $S$-factor for the whole range of energies under consideration. Intrinsically, in our calculations of twenty years' prescription [12], the behavior of the astrophysical $S$-factor of the radiative capture reaction p$^2$H $\rightarrow$ $^3$He$\gamma$ in the range from 10–20 to 150–200 keV, the value of which at these energies is determined generally by the $E$1 transition [17,19,20].

# 4 Radiative neutron capture on $^2$H in the potential cluster model

At first notice that the radiative neutron capture process on deuteron is considered, for example, in [68] in the frame of effective field theory. It was shown that the $M1$ transition gives the main contribution in the considering energy range 40–140 keV and it is possible to obtain a good agreement of the calculated total cross sections with their extrapolation from data base [69]. Furthermore, the possibility to describe experimental data on total cross sections of the radiative neutron capture on $^2$H at thermal (~1 eV), astrophysical (~1 keV), and low (~1 MeV) energies will be considered in the frame of the modified potential cluster model with forbidden states and their classification according to Young tableaux. The model and the numerical methods [21] of its realization developed here can describe correctly the behavior of the experimental cross sections at the energy range from 10 meV to 15 MeV.

## 4.1 Potential description of the n$^2$H elastic scattering

Furthermore, we will use the obtained above and in [7,8,20] p$^2$H potentials for consideration of the radiative neutron capture on $^2$H at low energies, using, at once, the same methods of calculations, which were checked for the p$^2$H system [7]. The parameters of the $^2S_{1/2}$ GS potential of $^3$H in the n$^2$H channel without Coulomb interaction were slightly improved for correct description of the binding energy of tritium, which is equal to -6.257233 MeV [49]. As a result, for the parameters of the potential of the form (3) was obtained [19]

$$V_{g.s.} = -41.4261655 \text{ MeV}, \quad \alpha_{g.s.} = 0.2 \text{ fm}^{-2}. \tag{36}$$

This potential reproduces the binding energy of $^3$H accurately, giving the value of -6.257233 MeV, obtained by the finite-difference method (FDM) [7], and it yields the charge and mass radii of 2.33 and 2.24 fm, respectively. The charge neutron radius equals zero, its mass radius equals proton radius of 0.8775(51) fm [34] and at the deuteron radius of 2.1424(21) fm [34]. Asymptotic constant (5) is equal to 2.04(1) at the interval of 5–15 fm. The AC error is formed by its averaging over the mentioned interval, but its values, obtained in different works, are given in [23] and are in the range of 1.82–2.21.

Let us note that the given here value of binding energy was determined at the accuracy of the FDM of $10^{-6}$ MeV, and, using the increased accuracy of 2 $10^{-9}$ for potential (36) it is possible to obtain more accurate value of -6.257233014 MeV. In addition, since deuteron has the radius more than tritium 1.755(86) fm [70], it can not be inside tritium in free, i.e., not deformed state, and the degree of its deformation, as it was shown in works [12], is equal near 30% [8].

The same conclusion was obtained in [51], where it was shown that the WF of deuteron located in tritium drops faster than the WF of deuteron in its free state. Thereby, the existence of the third particle, neutron in this case, leads to the deformation, i.e., compression of the deuteron cluster inside tritium nucleus. Approximately the same conclusion was done in the RGM calculations; the analysis of these results was done in work [2] and the usual estimation of the deuteron deformation is about 20%–40%.

Two-body variational method (VM) with the expansion of relative cluster motion WF by nonorthogonal Gaussian basis (10) and with the independent variation of all parameters [7,19] is used as an additional check to obtain the binding energy of $^3$H in the potential (36) for the n$^2$H channel. This variational method allows one to obtain the binding energy of -6.2572329999 MeV ≈ -6.257233000 MeV at the Gaussian basis having dimension $N$ = 10. The asymptotic constant $C_W$ of the variational WF, parameters of which are given in Table 5, remains at the level of 2.05(2) at distances of 6–20 fm that is not differ from the FDM value, and the residual errors are not more than $10^{-11}$ [7].

Table 5. Variational parameters and expansion coefficients of the bound state WF of $^3$H in the n$^2$H system. Normalization of function with these coefficients on an interval 0–25 fm is $N$ = 9.999999996433182E-001

| $i$ | $\beta_i$ | $C_i$ |
|---|---|---|
| 1 | 3.361218182141637E-001 | 1.231649877959069E-001 |
| 2 | 2.424705040532388E-002 | 1.492826524302106E-002 |
| 3 | 1.168704181683766E-002 | 1.190880013572610E-003 |
| 4 | 9.544908567362362E-002 | 1.304076551702031E-001 |
| 5 | 4.867951954385213E-002 | 5.868193953570694E-002 |
| 6 | 9.341901487408062E-001 | -2.155090483420204E-002 |
| 7 | 1.756025156195464E-001 | 1.814952898311890E-001 |
| 8 | 2.396705577261060E-001 | 6.944804259139825E-002 |
| 9 | 6.503621155681423E-001 | 1.564362603986158E-002 |
| 10 | 9.684977093058702E-001 | 1.709621746273126E-002 |

It is known that the variational energy decreases as the dimension of the basis increases and gives the upper limit of the true binding energy. At the same time the finite-difference energy increases as the size of steps decreases and the number of steps increases [7]. Therefore, it is possible to use the average value equals of 6.257233007(7) MeV for the n$^2$H system, obtained on the basis of two methods mentioned above, for the real binding energy in this potential. Thereby, we obtain that the accuracy of determination of the binding energy of this system in the listed above BS potential (36) obtained by two different methods (VM and FDM), on the basis of two different computer programs [7] is on the level of ±0.007 eV or ±7 meV [19].

## 4.2 The total cross sections of the radiative neutron capture on $^2$H

At first, we will show the working capacity of the modified potential cluster model used here, potentials obtained on the basis of the p$^2$H elastic scattering phase shifts and the corresponding GS potential of $^3$H on example of the photodisintegration of $^3$H into the n$^2$H channel. It was considered by us earlier in works [12] for more wide energy region, but less thoroughly. The results of these calculations at the energies of γ-quanta from 6.3 MeV to 10.5 MeV (we would remind you that the binding energy of $^3$H into the n$^2$H channel is equal to -6.257233 MeV) are shown in Fig. 4 by the solid line for the sum of the $E$1 and $M$1 cross sections with the given above p$^2$H potentials (see Table 1) with the Coulomb interaction switched off and with the GS potential (36).

The experimental data for total cross sections of the photodisintegration reaction of $^3$H into the n$^2$H channel for considered energies were taken from works: [71] – black triangles, [72] – open triangles.

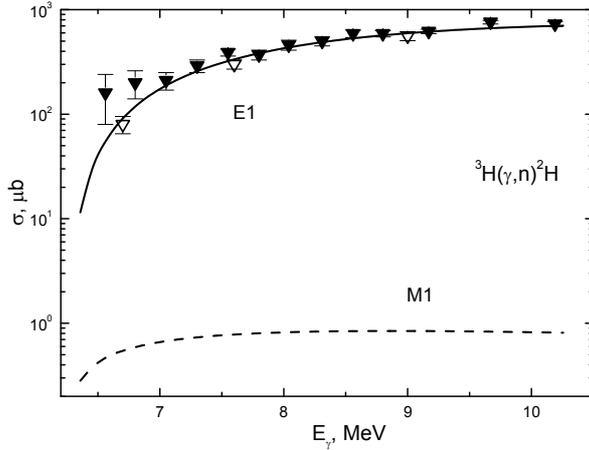

Fig. 4. The total photodisintegration cross sections of $^3$H into the n$^2$H channel. The experimental data: ▼ are from [71] and ∇ – [72]. Explanation for lines is in the text.

The contribution of the $M$1 process for disintegration of $^3$H in the ground $^2S_{1/2}$ state to the doublet $^2S_{1/2}$ wave of the n$^2$H scattering was shown by the dashed line in the bottom of the Fig. 4

1. $^2S_{1/2} \rightarrow {^2S_{1/2}}$,

which does not give appreciable contribution into the total cross sections of the reactions at these energies. Note once more attention to the fact that since different Young tableaux correspond to $^2S_{1/2}$ states of continuous and discrete spectra, they are matched to different interaction potentials. The cross sections of the considered process are caused exclusively by the $E$1 transition at the disintegration of the $^2S_{1/2}$ GS of $^3$H into the doublet $^2P$ scattering wave

2. $^2S_{1/2} \rightarrow {^2P_{3/2}} + {^2P_{1/2}}$.

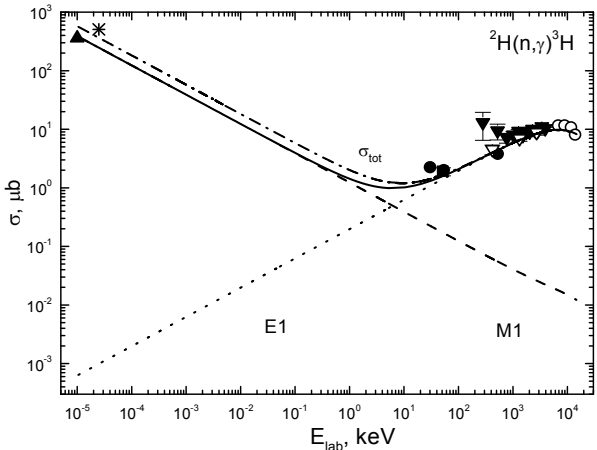

Fig. 5. The total cross section of the neutron radiative capture on $^2$H. The experimental data: ● are from [75], ○ – [76], ▲ – [73], * – [74], ■ – [77], recalculated for capture data: ▼ – [71] and ∇ – [72]. Explanations for lines are in the text.

Here, as opposed to our previous work [17], we will consider results for the neutron capture on $^2$H, when the negative sign of the neutron magnetic moment is taken into account [19]. The calculations of total cross sections of the radiative neutron capture on $^2$H at the energy range 10 meV – 15 MeV were done with the parameters of the p$^2$H nuclear potentials for the $^2S$ and $^2P$ scattering waves from Table 1 and for the GS (36) without Coulomb component. The results of calculations are shown in Fig. 5 by the dashed-dot line. It has happened that at the energies of 10 meV the calculated cross sections have the value slightly greater than the energies measured in other experiments [73], they more exactly agree with the data of [74] at 25 meV. The experimental data for total cross sections of the radiative neutron capture on $^2$H are taken from the works: [75] – points at energies 30, 55 and 530 keV, [76] – circles at 7–14 MeV, [73] – triangle at 0.01 eV, [74] – asterisk at 0.025 eV, [77] – square at 50 keV, and recalculated data of [71] marked in Fig. 4 by the reversed closed triangle (▼), and by the reversed open triangle (∇) [72].

The obtained earlier p$^2$H potential from Table 1 without Coulomb interaction for the $^2S$ scattering wave was used for calculations of the $M$1 transition No.1. But, let us note that the precision of results of different phase shift extractions obtained from the experimental data for the p$^2$H elastic scattering [44] reach 10%–20%, what is shown in Fig. 6 by points. Therefore, even the p$^2$H scattering potential, which phase shift is shown in Fig. 6 by the dashed line, is constructed on their basis with quite big errors, but here, we are considering the n$^2$H system, for what we did not succeeded in finding the results of phase shift analysis in the considered energy range.

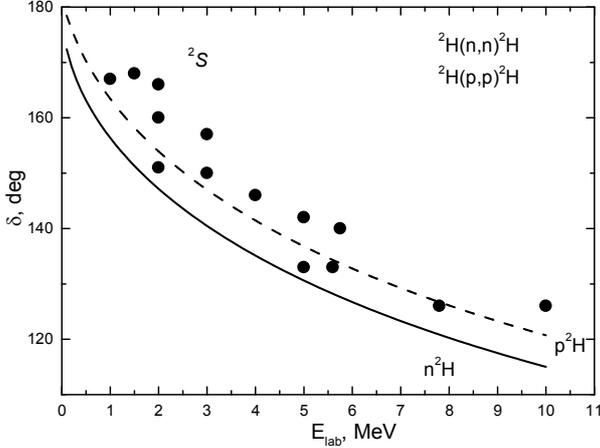

Fig. 6. The $^2S$ phase shifts of the p$^2$H (dashed line) and n$^2$H (solid line) elastic scattering. Points: • – the phase shifts obtained from the experimental data in [36]. The potential parameters are given in the text and in Table 3.1.

Therefore, furthermore we will consider the required changes, which will be necessary for the n$^2$H potential in the $^2S$ scattering wave, so that the result will be possible to describe the existent experimental data [73]. Consequently, the results for the summed for $E$1 with the transition $^2P_{3/2} + ^2P_{1/2} \rightarrow\, ^2S_{1/2}$ and $M$1 of the form No.1 total cross section of the radiative capture were shown in Fig. 5 by the solid line. The depth of the $^2S$ potential in the n$^2$H elastic scattering is not much smaller than for the p$^2$H system from Table 1 [19]

$$V_S = -52.0 \text{ MeV}, \quad \alpha_S = 0.2 \text{ fm}^{-2}. \qquad (37)$$

The scattering phase shift, obtained for this potential, is shown in Fig. 6 by the solid line. It is seen that the $^2S$ phase shift of the improved n$^2$H potential at low energies drops appreciably quicker than the analogous phase shift for the p$^2$H potential from Table 1. This fact, by-turn, affects the results of calculations of the total cross sections for the $M$1 process and, as it could be seen from Fig. 5, the usage of this potential allows to get a good description of the available data for total cross sections, even at the lowest energies 10 meV [73].

The results shown in Fig. 5 demonstrate prevalence of the $M$1 process at the energies lower 1 keV, this cross section is represented by the dashed line. The dotted line in Fig. 5 shows the contribution of the $E$1 transition. As it is seen from Fig. 5, the cross section of the $E$1 transition drops sharply and already at 0.1 keV it can be neglected. At the same time, this process at the range above 10 keV is dominant and absolutely determines the behavior of total cross sections, which allow us to describe the available experimental data at energies from 50–100 keV to 15 MeV.

Thereby, the change of parameters of the n$^2$H potential in the $^2S$ phase shift less than 5% relative to results from Table 1 allows us acceptably describe the available experimental data [73,74] at low energies. Such change of the parameters could be interpreted by the ambiguity of the existent p$^2$H phase shifts and their absence for the n$^2$H elastic scattering. Consequently, the using modified potential cluster model has allowed to reproduce correctly the experimental data for the total cross sections of the radiative neutron capture on $^2$H at the energy range, when energies at the edges of diapason differ from each other by more than nine orders, notably from 10$^{-5}$ keV to

1.5·10⁴ keV [17].

Since, the calculated cross section, which was shown in Fig. 5 by the solid line, is practically the straight line at energies from $10^{-5}$ to 0.1 keV, so it can be approximated by the simple function of the form

$$\sigma_{ap}(\mu b) = \frac{A}{\sqrt{E_n(\text{keV})}} . \qquad (38)$$

The value of the constant $A = 1.2314$ μb keV$^{1/2}$ was determined from the one point of the cross sections at the minimal energy, that equals of $10^{-5}$ keV. Furthermore, it is possible to consider the absolute value $M(E)$ of the relative deviation of the calculated theoretical cross section ($\sigma_{theor}$) and the approximation of this cross section ($\sigma_{ap}$) by function (38) in the range from $10^{-5}$ to 0.1 keV

$$M(E) = \left|[\sigma_{ap}(E) - \sigma_{theor}(E)]/\sigma_{theor}(E)\right| . \qquad (39)$$

It was found that at the energy range lower 100 eV this deviation does not exceed 1.0%. It is possible, evidently, to suppose that the shape of the energy dependence of the total cross section by energy (38) will be also preserved at lower energies. In this case, the estimation of the value of total cross section, for example at the energy 1 μeV, gives the value of 38.9 mb [19]. The analogous coefficient for the dashed-dot line in Fig. 5 is equal to $A = 1.8205$ μb keV$^{1/2}$, deviation of the calculation (39) and approximation (38) at 100 eV is at the level 1%, and the cross section value at 1 μeV equals approximately 57.6 mb [19]. Thus, the simple approximation of the calculated cross sections allows one easy use them in any other models and calculations.

## 5 Conclusion

Consequently, classification of cluster states according to Young tableaux allows one to determine that the potentials of the BSs and scattering states must be different at identical $L$, $S$ and $J$, but with the various tableaux $\{f\}$. Such potentials are constructed on the basis of description characteristics of the BSs and scattering phase shifts and allow correctly describe the available experimental data on the astrophysical $S$-factor of the proton radiative capture on ²H and total cross sections of the neutron capture on ²H with the formation of nuclei ³He and ³H.

Moreover, our calculations of the proton radiative capture on ²H for the $E$1 transition at the energy range down to 10 keV were carried out in work [12] in 1995, when the experimental data above 150–200 keV [45] was only known. It was found that these results describe well behavior of the $S$-factor from 10 keV to 200 keV, which was obtained in later measurements in works [46,47] in 1997 year at the energies down to 10 keV. Thus, the using modified potential cluster model with the classification of the orbital states according to Young tableaux allows one not only describe new experimental data, but, per se, predicts previously the behavior of the astrophysical $S$-factor of the proton radiative capture on ²H at the energy range from 10–20 to 150–200 keV.

The usage of practically the same potentials allows one to describe well the experimental data on the neutron radiative capture on $^2$H. Change of the *S* scattering wave potential by 5% allows one to match calculation results with one [73] or another [74] measurements at energy 10–25 meV. If other measurements of this cross section at thermal energies will be carried out in future, then owing to high dependence of the calculation cross section from the parameters of the *S* wave potential. Change of its parameters will be needed, evidently, in the range not more than 10%. Such changes, because there are no the n$^2$H phase shifts in the accessible for us literature at all, lie within the error band of the p$^2$H elastic scattering phase shifts, which reaches the value of 10–20%.

The stated procedure of construction nucleon-nucleus or cluster-nucleus potentials with the classification of states according Young tableaux allows one to obtain good results for description of the total cross sections and nuclear characteristics in 27 reactions of radiative capture at thermal and astrophysical energies [15-20]. Detailed list of all considered reactions is given in works [15-20] and books [7,12].

Thereby, it was shown that the MPCM completely allows not only well description of the available experimental data for total cross sections or astrophysical *S*-factors in the wide energy range, but even show the behavior of the *S*-factors at lowest energies. It is possible to obtain these results on the basis of methods that are appreciably easy than the RGM or the GCM [4] or, for example, hyperspherical harmonic method [78], which also lead to the good results at description of the radiative capture processes.

## Acknowledgments

This work was supported by the Grant Program No. 0151/GF2 of the Ministry of Education and Science of the Republic of Kazakhstan. We would like to express our thanks to Professor Strakovsky I. (GWU, USA) and Professor Blokhintsev L.D. (MSU, Russia) for discussion of certain questions touching upon in this work, and also to Professor Mukhamedzhanov A. (Texas Univ., USA) and Professor Yarmukhamedov R. for the detailed discussions of some questions of the work and for the provision of his results on the asymptotic normalization constants.